\documentclass[a4paper]{article}

\usepackage{INTERSPEECH2022}
\usepackage{subfigure}
\usepackage{caption}

\usepackage{multirow}
\usepackage{tablefootnote}
\usepackage[table,xcdraw]{xcolor}

\title{Investigation of Different Calibration Methods for Deep Speaker Embedding based Verification Systems}
\name{Galina Lavrentyeva$^{1,2}$, Sergey Novoselov$^{1,2}$, Andrey Shulipa$^1$, Marina Volkova$^{1,2}$, Aleksandr Kozlov$^2$}

\address{
  $^1$ITMO University, St. Petersburg, Russia\\
  $^2$STC-innovations Ltd., St. Petersburg, Russia
  }
\email{\{lavrentyeva, novoselov, shulipa, volkova, kozlov-a\}@speechpro.com}

\begin{document}

\maketitle
\begin{abstract}
Deep speaker embedding extractors have already become new state-of-the-art systems in the speaker verification field. However, the problem of verification score calibration for such systems often remains out of focus.
An irrelevant score calibration leads to serious issues, especially in the case of unknown acoustic conditions, even if we use a strong speaker verification system in terms of threshold-free metrics.

This paper presents an investigation over several methods of score calibration: a classical approach based on the logistic regression model; the recently presented magnitude estimation network MagnetO that uses activations from the pooling layer of the trained deep speaker extractor 
and generalization of such approach based on separate scale and offset prediction neural networks.

An additional focus of this research is to estimate the impact of score normalization on the calibration performance of the system. 

The obtained results demonstrate that there are no serious problems if in-domain development data are used for calibration tuning. Otherwise, a trade-off between good calibration performance and threshold-free system quality arises. In most cases using adaptive s-norm helps to stabilize score distributions and to improve system performance.

Meanwhile, some experiments demonstrate that novel approaches have their limits in score stabilization on several datasets.
\end{abstract}

\section{Introduction}
The revolutionary %TDNN (time delay neural network)  based  
x-vector  system proposed in \cite{Snyder} has switched the focus in the speaker recognition field to deep speaker embeddings. This system significantly outperformed conventional i-vector based system in terms of speaker recognition (SR) performance  and  hence  became a new baseline for text-independent SR tasks. 
The key feature of the proposed architecture is a statistics pooling layer designed to accumulate speaker information from the variable-length speech segment into a fixed-length vector which is then transformed into a low dimensional x-vector. Extracted speaker embeddings can be compared using for example conventional Probabilistic Linear Discriminant Analysis based backend model.
% смотрим на out-of-domain кейсы, когда системы разваливаются

The current state-of-the-art systems in the SR task are completely guided by the deep learning paradigm and are based on deep neural network speaker embeddings extractors. Previously frame level part of these extractors was based on TDNN (time delay neural network) blocks that contain only 5 convolutional layers with temporal context. Nowadays they use deeper ResNet architectures \cite{Gusev2020}, \cite{Zeinali2019}. The trend towards increasingly deep neural networks in image classification is also reflected in speaker recognition task \cite{Alam2020_resnet50}, \cite{Nagrani_resnet152}.
Moreover, these networks are trained using angular losses \cite{novoselov2018deep, Zeinali2019_BUT}, thus resulting in speaker embeddings that can be compared with cosine distance without the use of any trainable backend scoring model. % они дают близкие к PLDA результаты

Raw scores obtained both from the back-end model and by the cosine metric can be effectively utilized for solving speaker recognition tasks in the concrete acoustic domain by adjusting the decision threshold. However, in most cases, the mismatches in training and testing conditions lead to scores distributions scaling and shifting for different domains and signal quality (SNR, duration). Such scores instability impairs SR systems in real applications. In order to deal with this problem different calibration strategies \cite{PLDA_Ferrer2020}, ~\cite{Calibration_Shulipa2016} and compensation techniques \cite{QE_Lavrentyeva2020}, \cite{Magneto_Garcia-Romero2020} were implemented.

%Authors of \cite{PLDA_Ferrer2020} proposed a new backend in a functional form that mimics the standard PLDA. They report that this backend takes into account the variety of acoustic conditions and can result in well-calibrated scores without the need for further calibration.

This paper presents an investigation over several methods of score calibration: a classical approach based on the logistic regression model; recently presented magnitude estimation network MagnetO \cite{Magneto_Garcia-Romero2020} that uses activations from the pooling layer of the trained deep speaker extractor and generalization of such approach based on separate scale and offset prediction neural networks developed during the study.

% An additional focus of this research is the impact estimation of score normalization \cite{CVDFKCL2017} on the calibration performance of the system.

%Adding adapted top s-normalization to the calibration step helps to improve the quality of the final system.
Based on our experience in different speaker recognition challenges with cross condition evaluations \cite{avdeeva2021stc} we were also interested in the importance of score normalization in all of these approaches. Due to this, we additionally considered the use of adapted s-normalization \cite{CVDFKCL2017} after the calibration to analyze its effectiveness.

Another focus in this investigation is devoted to the question if additional information or regularisation during training neural network based calibration approaches can help to improve the final performance of the SR system.
During training the neural network based calibration back-end several experiments were conducted where additional input with utterance duration and regularisation techniques based on the domain information were implemented.

%Let us start from brief overview of the calibration methods we investigated. 

\section{Problem overview}

The mismatches in training, testing, and enrollment conditions can lead to scores distributions scaling and shifting. Thus the scores produced by the model, no matter if PLDA backend or cosine metric is used, are not well calibrated or not calibrated at all. And additional calibration step that transforms the scores to proper log-likelihood ratios is demanded.

The standard calibration recipe supposes to apply an affine transformation to the raw scores to compensate its shift and scale for mismatched data conditions.
The classical approach implies using linear logistic regression to train the parameters of this affine transformation, using binary cross-entropy ~\cite{Brummer2013}. Let $s_{ij}$ and $l_{ij}$ be the raw and calibrated scores for $i$ and $j$ utterances comparison, then calibration can be processed by
\begin{equation}
\label{eq:cal}
    l_{ij} = \alpha s_{ij} + \beta
\end{equation}

where $\alpha, \beta$ are scaling and offset calibration parameters. These parameters are usually trained on a calibration training database using prior-weighted binary cross entropy loss.
\begin{equation}
    C_{\pi} = \dfrac{\pi}{|T|} \sum_{ij \in T} \log(1 + e^{-p_{ij}}) + \dfrac{1 - \pi}{|N|} \sum_{ij \in N} \log(1 + e^{p_{ij}})
\end{equation}
where $\pi$ is the prior probability for a target trial (operating point), T is a set of target trials and N is a set of non-target trials and $p_{ij} = l_{ij} + \log(\pi/1-\pi)$.
Another option is to use more general version of this loss with operating point being integrated out. It is called cost of the likelihood ratio (Cllr) and indicates score calibration across all operating points along a detection error tradeoff (DET) curve \cite{Brummer2014}. This metric is usually used for calibration performance measurements. According to our observations there is no big difference in performance of the systems trained with these losses.
Many papers proposed to use duration, domain labels or other additional information to improve calibration capability \cite{Nautsch}.

\subsection{Condition-aware calibration }
Our work about the blind quality estimation of the speech signal ~\cite{QE_Lavrentyeva2020} confirms that additional information about enrollment and test conditions obtained by quality assessments can compensate speech condition scores shifting in speaker verification task. According to the results presented in ~\cite{QE_Lavrentyeva2020} Quality Measure Functions based on signal-to-noise-ratio and reverberation time allow to improve the SR system performance on VOiCES eval protocol, where SR systems demonstrated performance degradation compared to results obtained for the development protocol \cite{Gusev2020}.

\begin{figure}[]
  \centering
  \captionsetup[subfigure]{labelformat=empty}
  \subfigure[]{\includegraphics[width=0.46\linewidth]{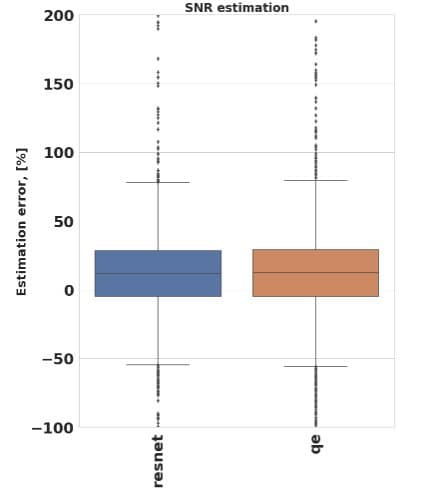}} 
  \subfigure[]{\includegraphics[width=0.46\linewidth, trim =0 0 0 0]{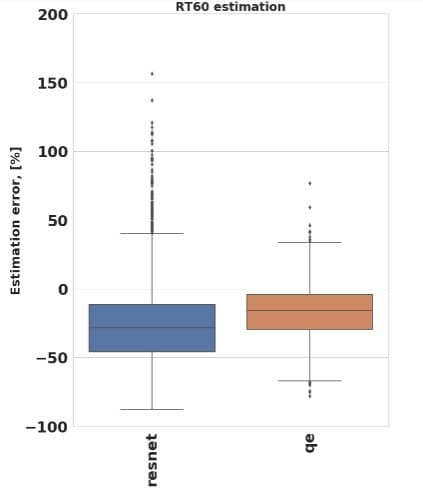}} 
  
  \caption{SNR(a) and RT60(b) estimations errors for the ACE eval dataset for teacher (qe) and student (resnet) models}
  \label{fig:boxplots}
\end{figure}

Our additional experiments confirm that statistic pooling layer of the deep speaker embedding extractor also contains information about acoustic conditions. In these experiments, the fixed ResNet system trained for speaker recognition task was used to extract the activations of its pooling layer. These values were then used as an input to train simple neural network model with several fully connected layers to predict SNR and RT60 of the audio samples. This model was successfully trained using the Quality Estimation (QE) network from ~\cite{QE_Lavrentyeva2020} as a teacher.
Prediction errors of the teacher (QE) and student (ResNet) models are very close. Box-plots in Figure \ref{fig:boxplots} displays the median and dispersion of the SNR and RT60 estimations errors for the ACE eval dataset \cite{ACE}.

Results from these experiments in addition to the impressive results of the mentioned above papers lead us to a conclusion that post-pooling activations contain specific information for the calibration process and can be used for condition-aware calibration.

\subsubsection{MagNet-O}
The above conclusions correspond to the results of \cite{Magneto_Garcia-Romero2020} where authors proposed a magnitude estimation network that uses extracted x-vectors (statistic pooling activations of the TDNN system preliminary trained with angular softmax) to predict calibrated scores. This network maps speaker embedding to its magnitude, which plays the role of a scale factor in a classic calibration approach based on the linear regression model. It is trained using pairs of target and impostor trials with prior-weighted binary cross-entropy loss. 
%Since this approach is not a linear transformation of the scores it not only calibrates the system's scores but also improves the discriminating ability of the system.
Since this approach is not a linear transformation of the scores it was mainly proposed by the authors to improve the discriminating ability of the system.

\subsubsection{Generalized model}

Inspired by the results obtained in \cite{Magneto_Garcia-Romero2020} we consider a generalized neural network based approach for condition-aware speaker embeddings calibration. The calibration neural network is trained directly for the speaker verification task, using binary classification for target and non-target trials. 
Similar to the work mentioned above, this approach uses post-pooling activation to predict the scale and offset parameters used to present scores as log-likelihood-ratios. However, it can be considered as a more general approach as it does not map its input to the magnitude for each embedding but produces scale and offset parameters for each trial. This approach allows to use the enrollment and test utterances duration (or any other specific information) as an additional input and forces the neural network to take into account the differences in its conditions. Moreover, we use two different neural networks for scale and offset parameters. 

In our experiments, we fixed one of our ResNet-based embedding extractors and compare several approaches for its scores calibration. 
We consider standard calibration via linear logistic regression as a baseline system. We also investigate MagNetO approach with our embedding extractor and compare it with our modifications on the public datasets from NIST SREs (NIST 2016, NIST 2019), Voices Challenge, and our private dataset STC\_calls for cross channel test.

% %Adding adapted top s-normalization to the calibration step helps to improve the quality of the final system.
% Based on our experience in different speaker recognition challenges with cross condition evaluations \cite{avdeeva2021stc} we were also interested in the importance of score normalization in all of these approaches. Due to this, we additionally considered the use of adapted s-normalization after the calibration to analyze its effectiveness.

% Another focus in this investigation is devoted to the question if additional information or regularisation during training neural network based calibration approaches can help to improve the final performance of the SR system.
% During training the neural network based calibration back-end we conduct several experiments with additional input with utterance duration and regularisation techniques based on the domain information.

%Thus the question of training robust calibration neural networks is still open.
%The analysis was performed for common speaker recognition benchmarks with varying acoustic conditions, languages, and channels: NIST SRE 2016, NIST SRE 2019, VOiCES, our Russian in-house dataset STC\_calls.  

% \section{Preliminaries}

\section{Embedding Extractor}
During our experiments, we used a deep speaker embedding extractor with residual network architecture named ResNet-101 \cite{He2016}. %Table \ref{tab:arch} describes the architecture we used. 
The key block of ResNet-101 is ResNetBlock. It consists of three convolutional layers in ResNetBlock with 1×1, 3×3, and 1×1 masks. ReLU activation follows each convolutional layer, and Maxout activation is used for embedding extraction. We apply the batch normalization technique to stabilize and speed up network convergence.
This extractor was trained on 64 Mel filter bank log-energies with UNet VAD from \cite{Gusev2020}.

\begin{table*}[]
\caption{Performance of the considered calibration approaches in terms of EER, minDCF and actDCF for $P_{tar} = 0.05$}
\label{tab:eer}
\resizebox{\textwidth}{!}{
\begin{tabular}{|l|l|l|l|l|l|l|l|l|l|l|l|l|}
\hline
\cellcolor[HTML]{FFFFC7} & \multicolumn{3}{c|}{\cellcolor[HTML]{FFFFC7}Matched dev-test conditions} & \multicolumn{9}{c|}{\cellcolor[HTML]{FFFFC7}Mismatched dev-test conditions} \\ \cline{2-13} 
\cellcolor[HTML]{FFFFC7} & \multicolumn{3}{c|}{\cellcolor[HTML]{FFFFC7}NIST2019 eval} & \multicolumn{3}{c|}{\cellcolor[HTML]{FFFFC7}NIST2016 eval} & \multicolumn{3}{c|}{\cellcolor[HTML]{FFFFC7}VoiCES eval} & \multicolumn{3}{c|}{\cellcolor[HTML]{FFFFC7}STC\_calls cc} \\ \cline{2-13} 
\rowcolor[HTML]{FFFFC7} 
\multicolumn{1}{|c|}{\multirow{-2}{*}{\cellcolor[HTML]{FFFFC7}Systems}} & EER & $C_{min}^{0.05}$ & $C_{act}^{0.05}$ & EER & $C_{min}^{0.05}$ & $C_{act}^{0.05}$ & EER & $C_{min}^{0.05}$ & $C_{act}^{0.05}$ & EER & $C_{min}^{0.05}$ & $C_{act}^{0.05}$ \\ \hline \hline

Baseline & \multicolumn{1}{l|}{2.779} & \multicolumn{1}{l|}{0.162} & 0.185 & \multicolumn{1}{l|}{6.844} & \multicolumn{1}{l|}{0.408} & \multicolumn{1}{l|}{0.577} & \multicolumn{1}{l|}{5.933} & \multicolumn{1}{l|}{0.219} & \multicolumn{1}{l|}{0.54} & \multicolumn{1}{l|}{12.478} & \multicolumn{1}{l|}{\textbf{0.604}} & 0.975 \\ \hline
Baseline + snorm & \multicolumn{1}{l|}{2.931} & \multicolumn{1}{l|}{0.169} & 0.195 & \multicolumn{1}{l|}{5.668} & \multicolumn{1}{l|}{\textbf{0.256}} & \multicolumn{1}{l|}{\textbf{0.257}} & \multicolumn{1}{l|}{6.331} & \multicolumn{1}{l|}{0.229} & \multicolumn{1}{l|}{0.245} & \multicolumn{1}{l|}{12.907} & \multicolumn{1}{l|}{0.609} & 0.737 \\ \hline
MagNetO & \multicolumn{1}{l|}{2.648} & \multicolumn{1}{l|}{0.16} & 0.165 & \multicolumn{1}{l|}{6.672} & \multicolumn{1}{l|}{0.447} & \multicolumn{1}{l|}{2.418} & \multicolumn{1}{l|}{5.44} & \multicolumn{1}{l|}{0.21} & \multicolumn{1}{l|}{0.242} & \multicolumn{1}{l|}{11.957} & \multicolumn{1}{l|}{0.631} & 0.723 \\ \hline
MagNetO + dur & \multicolumn{1}{l|}{2.704} & \multicolumn{1}{l|}{0.158} & 0.165 & \multicolumn{1}{l|}{6.478} & \multicolumn{1}{l|}{0.402} & \multicolumn{1}{l|}{2.114} & \multicolumn{1}{l|}{5.607} & \multicolumn{1}{l|}{0.216} & \multicolumn{1}{l|}{0.256} & \multicolumn{1}{l|}{12.067} & \multicolumn{1}{l|}{0.631} & 0.743 \\ \hline
MagNetO + dur + snorm & \multicolumn{1}{l|}{2.681} & \multicolumn{1}{l|}{0.157} & \textbf{0.161} & \multicolumn{1}{l|}{5.475} & \multicolumn{1}{l|}{0.332} & \multicolumn{1}{l|}{0.512} & \multicolumn{1}{l|}{6.52} & \multicolumn{1}{l|}{0.236} & \multicolumn{1}{l|}{0.363} & \multicolumn{1}{l|}{12.89} & \multicolumn{1}{l|}{0.617} & 0.795 \\ \hline
MagNetO + dur + stdloss & \multicolumn{1}{l|}{2.802} & \multicolumn{1}{l|}{0.163} & 0.167 & \multicolumn{1}{l|}{6.865} & \multicolumn{1}{l|}{0.406} & \multicolumn{1}{l|}{0.944} & \multicolumn{1}{l|}{5.962} & \multicolumn{1}{l|}{0.22} & \multicolumn{1}{l|}{0.424} & \multicolumn{1}{l|}{12.51} & \multicolumn{1}{l|}{0.607} & 0.935 \\ \hline
MagNetO + dur + snorm + stdloss & \multicolumn{1}{l|}{2.931} & \multicolumn{1}{l|}{0.169} & 0.193 & \multicolumn{1}{l|}{5.668} & \multicolumn{1}{l|}{\textbf{0.256}} & \multicolumn{1}{l|}{\textbf{0.257}} & \multicolumn{1}{l|}{6.331} & \multicolumn{1}{l|}{0.229} & \multicolumn{1}{l|}{0.244} & \multicolumn{1}{l|}{12.907} & \multicolumn{1}{l|}{0.609} & 0.74 \\ \hline
SONet & \multicolumn{1}{l|}{2.891} & \multicolumn{1}{l|}{0.165} & 0.168 & \multicolumn{1}{l|}{6.588} & \multicolumn{1}{l|}{0.399} & \multicolumn{1}{l|}{2.042} & \multicolumn{1}{l|}{4.512} & \multicolumn{1}{l|}{0.196} & \multicolumn{1}{l|}{0.261} & \multicolumn{1}{l|}{\textbf{11.5}} & \multicolumn{1}{l|}{0.609} & 0.746 \\ \hline
SONet + dur & \multicolumn{1}{l|}{2.697} & \multicolumn{1}{l|}{0.158} & 0.168 & \multicolumn{1}{l|}{6.568} & \multicolumn{1}{l|}{0.387} & \multicolumn{1}{l|}{1.699} & \multicolumn{1}{l|}{\textbf{4.383}} & \multicolumn{1}{l|}{\textbf{0.193}} & \multicolumn{1}{l|}{0.295} & \multicolumn{1}{l|}{11.787} & \multicolumn{1}{l|}{0.638} & 0.782 \\ \hline
SONet + dur + snorm & \multicolumn{1}{l|}{\textbf{2.647}} & \multicolumn{1}{l|}{\textbf{0.154}} & 0.167 & \multicolumn{1}{l|}{\textbf{5.466}} & \multicolumn{1}{l|}{0.292} & \multicolumn{1}{l|}{0.883} & \multicolumn{1}{l|}{4.517} & \multicolumn{1}{l|}{0.199} & \multicolumn{1}{l|}{0.236} & \multicolumn{1}{l|}{11.582} & \multicolumn{1}{l|}{0.623} & \textbf{0.713} \\ \hline
SONet + dur + stdloss & \multicolumn{1}{l|}{2.795} & \multicolumn{1}{l|}{0.163} & 0.181 & \multicolumn{1}{l|}{6.813} & \multicolumn{1}{l|}{0.399} & \multicolumn{1}{l|}{1.132} & \multicolumn{1}{l|}{5.918} & \multicolumn{1}{l|}{0.219} & \multicolumn{1}{l|}{0.381} & \multicolumn{1}{l|}{12.494} & \multicolumn{1}{l|}{0.606} & 0.91 \\ \hline
SONet + dur + snorm + stdloss & \multicolumn{1}{l|}{2.938} & \multicolumn{1}{l|}{0.169} & 0.182 & \multicolumn{1}{l|}{5.676} & \multicolumn{1}{l|}{\textbf{0.256}} & \multicolumn{1}{l|}{\textbf{0.257}} & \multicolumn{1}{l|}{6.344} & \multicolumn{1}{l|}{0.23} & \multicolumn{1}{l|}{\textbf{0.234}} & \multicolumn{1}{l|}{12.927} & \multicolumn{1}{l|}{0.609} & 0.767 \\ \hline
\end{tabular}
}
\end{table*}

This neural network was firstly pre-trained on the train dataset with 4.5-sec utterance duration and after that fine-tuned on utterances with a longer duration (12 sec). In both cases, the training process was performed with additive angular margin softmax loss. The described network was fixed and used further in all conducted experiments for embedding extraction from the second from the last dense layer (Dense1).

% \begin{figure*}[]
%   \centering
%   \captionsetup[subfigure]{labelformat=empty}
%   \subfigure[]{\includegraphics[width=0.48\linewidth]{images/nist2016_all.png}} 
%   \subfigure[]{\includegraphics[width=0.48\linewidth]{images/nist2019_all.png}} 
%   \subfigure[]{\includegraphics[width=0.48\linewidth]{images/voices_all.png}}
%   \subfigure[]{\includegraphics[width=0.48\linewidth]{images/stc_cc_all.png}} 
%   \caption{Performance of the considered calibration approaches in terms minDCF and actDCF for $P_{tar} = 0.05$ and $P_{tar} = 0.01$}
%   \label{fig:dcf}
% \end{figure*}

\section{Calibration}
%Scores of the extracted speaker embeddings are further calibrated by one of the considered approaches or its modifications.
Extracted embeddings are further used for score estimation for each trial. As it was mentioned before current state-of-the-art models trained with angular losses mostly use the cosine similarity metric since it shows comparable performance to PLDA backends. That is the reason it was used in our experiments as well.

As a baseline we use a standard approach with logistic regression described earlier. The baseline was trained using CLLR loss on NIST SRE 2018 dev dataset only.

In the second step, the aim was to reproduce the MagNetO approach with our ResNet-101 extractor.
Speaker embeddings are used here as an input to the neural network with MagNetO-2-like topology. It consists of 3 affine layers and contains 2.8M parameters in total. We changed ReLU to PReLU activations as it demonstrated better performance and we did not use any data augmentation in order to speed up the experiment.

MagNetO network is trained on pairs of target and impostor trials with prior-weighted binary cross-entropy loss. However, this approach estimates the magnitude of each embedding independently and does not consider the embedding of the other trial used in the scoring procedure. This lead us to a more general approach we called SONet (Scale and Offset networks) by analogy with MagNetO. Firstly, unlike MagNetO, SONet takes as an input both enrollment and test statistics pooling outputs concatenated into one. This allows training calibration more accurately with respect to possible mismatches in acoustic conditions.
Secondly, Offset is predicted by a different neural network. We consider it as a generalization of the previous approach.
In our experiments, MagNetO and SONet were trained with Cllr loss, since it shows similar results to prior-weighted binary cross-entropy loss used in the original paper.

Since the main goal was to evaluate how well the systems can be calibrated for all acoustic conditions.
To perform an accurate comparison for all the considered systems we have to put them in the same circumstances. This means that their decision thresholds have to be tuned on the same calibration set.
For this purpose in our experiments, all MagNetO and SONet based systems were additionally tuned on NIST2018 dev dataset using the classical calibration approach.

\subsection{Modifications}
\subsubsection{Duration}
First of all, we propose that adding information about the duration of the utterances as an input to the deep neural network should help to improve the calibration. According to \cite{Calibration_Shulipa2016} we used the logarithm of duration.
\subsubsection{Score normalization}
Another modification we implemented was score normalization. It is usually used to reduce the variability of trial scores. As a result, it leads to better calibration and improves system performance. In the proposed systems we apply the adaptive symmetric normalization technique (S-norm) as presented in \cite{Matejka2017}. This approach implies using an adaptive cohort for each trial, selected from normalization dataset as the X utterances closest to the test $t$ and enrollment $e$. If $S_e^{top}$ and $S_t^{top}$ are sets of X top scores, chosen for enrollment and test utterances $e$ and $t$, $s(e,t)$ is a raw score, then normalized score can be computed as follow:
\begin{equation}
\label{eq:s}
    s(e,t)_{norm} = \dfrac{1}{2}\left(\dfrac{s(e,t)-\mu(S_e^{top})}{\sigma(S_e^{top})}+\dfrac{s(e,t)-\mu(S_t^{top})}{\sigma(S_t^{top})}\right)
\end{equation}
To perform the adapted s-normalization in our experiments we used BABEL speech dataset as a normalization set. It contains 7376 utterances from different channels in several languages. We selected 200 top closest scores as described above for each trial to get a normalized score accordingly to the equation \ref{eq:s}.    
\subsubsection{Regularization}
According to the \cite{Magneto_Garcia-Romero2020} neural network based calibration trained with prior-weighted binary cross-entropy loss presents "non-monotonic mapping of the scores, and therefore, it can improve the speaker discrimination". In other words, a calibration neural network is partially trained to solve speaker recognition tasks.

During our investigation, we explored the training loss improvement by enforcing the "calibration" properties of the system. The main purpose was to make scale and offset factors speaker-independent but remain acoustic dependent. This can be achieved by acoustic domain-dependent batch training procedure with standard deviation loss regularization of the scale and offset factors within the batch. This std loss regularization is used in combination with Cllr loss.

\section{Experimental setup}
\subsection{Training dataset}
In all our experiments we used one fixed training dataset consisting of telephone and microphone data from various public and private datasets. It includes Switchboard2 Phases 1, 2, and 3, Switchboard Cellular, data from NIST SREs from 2004 through 2010 and 2019, VoxCeleb 1,2 \cite{Voxceleb, Voxceleb2} and an extended version of Russian speech subcorpus named RusTelecom v2. RusTelecom is a private Russian speech corpus of telephone speech, collected by call-centers in Russia. 

In order to increase the amount and diversity of the training data, we added augmented data produced by the standard Kaldi augmentation recipe (reverberation, babble, music, and noise) using the freely available MUSAN and simulated Room Impulse Response (RIR) datasets. In total, training dataset contains 1,679,541 recordings from 33,466 speakers.
% сократить

\subsection{Test datasets}
Our experimental setup includes evaluation on the most popular datasets NIST2016 eval, NIST2019 eval, VOiCES eval, and private STC\_calls subset, mainly used to estimate calibration performance in the challenging scenario with different domains of enrollment and test.

% The base IVR\_calls is a proprietary base of STC company. It includes voices of 1000 speakers, collected in various noise conditions. The telephone part of the base represents the interaction with the IVR-system with the following scenario: pronouncing 24 standard short phrases like «name and surname», «balance inquiry» etc., and 4 long stories on a given topic. 
% During phone calls, the voice of the speaker was recorded in parallel on various microphone devices. In this paper, we used only cross-channel protocol with a 30-sec long telephone enrolment and 5-sec long microphone test, collected on a far-field microphone array. This subset is denoted as STC\_calls cc.

The base STC\_calls is a private STC base. It contains voices of 1000 speakers, collected in various noise conditions simultaneously on the telephone and several microphone devices. It represents the interaction with the IVR system in 2 predefined scenarios: for text-dependent and text-independent cases.% with the following scenario: pronouncing 24 standard short phrases like «name and surname», «balance inquiry» etc., and 4 long stories on a given topic. 
In our experiments, we used only cross-channel protocol with 30-sec long telephone enrolment and 5-sec long microphone test, collected on a far-field microphone array. This subset is denoted as STC\_calls.

Since the final systems tuning of scale and offset for all the systems was performed using NIST2018 dev set, here only for NIST2019 eval the development set contains in-domain data, other test sets represent mismatched dev-test acoustic conditions: NIST2016 eval set contains utterances of different language, VoiCES eval set represents a different channel, and STC\_calls cc includes both different language and channel.

\subsection{Evaluation metrics}
We evaluate speaker recognition system performance in terms of Equal Error Rates (EER) and minimum detection cost functions with  $P_{tar}=0.05$: $C_{min}^{0.05}$ \cite{Calibration_Shulipa2016, evalplan2021nist}. To estimate calibration performance of the systems at this point we use conventional actual detection cost function $C_{act}^{0.05}$ \cite{evalplan2021nist}.

\section{Experimental results and discussions}
According to the results, presented in Table \ref{tab:eer}, baseline system demonstrates good calibration for in-domain data (NIST2019 eval) and bad calibration for other conditions. MagnetO allows increasing the quality of SR system in terms of EER, while SONet provides an additional performance gap, achieving 4.5\% EER on the VOiCES eval set. 
On the one hand, it should be noted that these are impressive results compared to top task-oriented systems from \cite{Gusev2020}.
On the other hand, the original MagNeto without any modifications has its limits in terms of its calibration ability for varying acoustic conditions. Being its generalization, SONet seems to have the same problems, even though it can achieve better performance, for example on Voice eval set.

In case the development dataset acoustic conditions match those of the test set, MagNetO and SONet based approaches can help to improve both the quality of the final systems in terms of EER and its calibration ability. Which means that there is no problem with the in-domain calibration task. And results for NIST2019 prove that.

Score normalization brings a stable gain in terms of both SR and calibration quality, even with out-of-domain s-norm cohort (VOiCES, STC\_calls cc). The baseline system with this modification show extremely good results in terms of minDCF.

Duration-aware versions of MagNetO and SONet demonstrate benefits in some cases (NIST2016, VOiCES) compared to their original versions, however, it does not seems to be a common solution for all tests with mismatches conditions.

An interesting observation is that weighted cross-entropy and Cllr loss used for training here have complex properties. There are two options to decrease calibration training losses. One of them is to directly improve speaker discrimination properties of the verification system. From this point of view, the calibration operation in eq.\ref{eq:cal} can be considered as a ``fusion'' model of origin raw score and offset ``score''. The weights of this fusion are determined by scale factor $\alpha$. Since the stat-pooling layer output is rich in speaker information, scale and offset factors in this scenario should be considered to be speaker-dependent. Our experiments show that training the system with Cllr or weighted cross-entropy loss without any regularization components leads to better discriminative properties of the SONet system compared to the baseline system. And as an opposite, it has not so good calibration or stabilization characteristics. %It should be noted surprisingly impressive improvements were obtained for VOiCES evaluation protocols in spite of the fact that it has out-of-domain (OOD) nature for development set and adaptive s-norm set. 
%Comparative analysis on Figure \ref{fig:dcf}, 
Experimental results, presented in Table \ref{tab:eer}, confirms that proposed std loss regularization combined with score normalization allows to achieve low actual costs for both MagNetO and SONet based systems on all considered test benchmarks and at the same time it slightly degrades the quality of these systems in terms of EER.

\section{Conclusions}
The comparative analysis of classical and new recently proposed solutions for score calibration was carried out in this paper. Additionally to this, a generalized score calibration model was described with several modification ideas, that can help to improve the final system performance and/or its robustness on out-of-domain data.
Experimental results, obtained here confirm that the statistic pooling layer of deep speaker embeddings extractor provides useful information to perform SR system calibration for a variety of acoustic conditions.  
We found out and verified that such calibration systems could be trained using the benefits of adaptive s-norm scores normalization. The duration-aware calibration systems could be slightly better than the systems without %enroll and test 
duration inputs. Such kind of ``tricks'' help to improve scores stabilization and enforce the robustness of the verification system to different acoustics conditions.
For strong calibration performance, regularization techniques like offered std loss component can be used.

An overall recommendation, based on the obtained experimental results, is that if a system robust to varying acoustic conditions is required it is better to use the classical calibration approach for raw scores after score normalization.
However, if there are in-domain data and the system needs to be tuned for these specific conditions, SONet based approach with the proposed modifications can help and calibration of the final system will not degrade.

Therefore, the question of the robust calibration of speaker recognition systems based on neural networks remains open.

%It should be noted surprisingly impressive improvements obtained for VOiCES evaluation protocols in spite of the fact that it has out-of-domain (OOD) nature for development set and adaptive s-norm set. 

%S-norm and duration aware versions of MagNetO and SONet demonstrates benefits in both quality and calibration performance in contrast to their original versions.
\vfill\pagebreak
\bibliographystyle{IEEEtran}


\end{document}